\documentstyle[11pt,fullpage,doublespace,epsf]{article}
 \DeclareMathSizes{11}{19}{13}{9}   % For size 11 text

\makeatletter
\@addtoreset{equation}{section}
\makeatother

\begin{document}

  \title{Toward the Graphics Turing Scale\\
 on a Blue Gene Supercomputer}
\author{Michael McGuigan\\Brookhaven National Laboratory\\Upton NY 11973\\mcguigan@bnl.gov}
\date{}
\maketitle

\begin{abstract}We investigate raytracing performance that can be achieved on a class of Blue Gene supercomputers. We measure a 822 times speedup over a Pentium IV on a 6144 processor Blue Gene/L. We measure the computational performance as a function of number of processors and problem size to determine the scaling performance of the raytracing calculation on the Blue Gene. We find nontrivial scaling behavior at large number of processors. We discuss applications of this technology to scientific visualization with advanced lighting and high resolution. We utilize three racks of a Blue Gene/L in our calculations which is less than three percent of the the capacity of the worlds largest Blue Gene computer.
\end{abstract}

\section{Introduction}

The Graphics Turing scale can be defined by the computational ability to produce photorealistic imagery at 30 frames per second \cite{mcguigan}. Like the traditional Turing scale indistinguishability is the  main criterion \cite{Turing}. In pursuing this computational scale one can use whatever computational approach one likes to produce the imagery. One can use special hardware for performing near real time raytracing \cite{Woop}. One can harness the power of recent programmable graphics processing units (gpus) to render the scenery \cite{Hanrahan} \cite{Popov}. One can use a large graphics grid or render farm to create the image as is currently done in most computer animation studios. In this paper we use part of a 103 Teraflop/s (103 Trillion Floating point operations per second) Blue Gene/L (BGL) supercomputer to render computer imagery. We then study the scaling behavior as one increases the number of processors applied to the graphics computation. The Blue Gene/L supercomputer we used consists of 18 racks, each rack has 1024 nodes and each node has two processors. Each processor is a PowerPC 400 700MHz rated at (2.8 Gflop/s) (2.8 Billion floating point operations per second). A description of the computer is listed in Table 1.

\begin{table}[h]
\begin{center}
\begin{tabular}{ |c|c|c|c|c|c|c|} \hline
 Rank   & Facility & Racks & Nodes & Processors & Peak Speed &Memory\\ \hline
10 &BNL & 18 racks & 18432 & 36864 &  103 Teraflop/s &18.4TB\\ \hline
\end{tabular}
\end{center}
\caption{Characteristics of the Blue Gene/L supercomputer used for the graphics computations in this paper. $1$ Terasflop/s $= 10^{12}$ floating point operations per second. $1$ TB $=1000$ Gigabytes. }
\label{tab1}
\end{table}

Reaching the Graphics Turing scale has applications to virtual reality, simulation, animation and visualization \cite{Kreuger}. As the Blue Gene computer is used for scientific research we were mainly interested in the application to scientific visualization and in particular parallel visualization \cite{Ahrens} \cite{Tomov1}.

This paper is organized as follows. In section 2 we discuss our approach describing the method, scientific model and software used.  We study strong scaling of the graphics application at high resolution (17.6 Megapixels) with advanced lighting. In section 3 we study weak scaling of the graphics computation or what happens when one increases the computational cost at the same time one increases the number of processors. In section 4 we state the main conclusions.

 \section{Strong scaling}

While several excellent renderers are available for example povray, rayshade and renderman we chose tachyon \cite{Stone} because it was designed from the outset for speed and for portability to a large number of advanced computer systems. For scientific visualization tachyon can be used to render data sets from the molecular visualization software VMD \cite{VMD} as well as data from the mathematical software SAGE \cite{sage}. Tachyon is also part of the SpecIntMPI2007 benchmarking suite used to measure the speed of parallel computing platforms \cite{mpi}.

We used tachyon 0.98 version dated 3/19/2007. The code is written in the C programming language and uses the MPI message passing interface for parallel programming. Compiling the C source code for tachyon was straightforward on the Blue Gene with the use of the cross compiler between the front end node and the compute nodes. The code was compiled using the double precision option for all calculations. We ran the code in VN mode in which both processors of each node of the BGL are used for computation.

For the scientific data set we chose the VMD example scene downloadable from
\cite{VMD}. This scene contains 26318 objects and the 1e79 Atp phosphorylase molecular model is depicted in Figure 1. Important criteria that determine the computational workload for this data set is the resolution and the complexity of the lighting model. 

At 1.1 Megapixel  resolution and ambient occlusion lighting the tachyon rendering command we used was \cite{VMD}:

{\tt \small \it
 tachyon -res 965 1137 -trans\_vmd -rescale\_lights 0.3 -add\_skylight 1.0 1e79ao.dat
}

At 17.6 Megapixel resolution and ambient occlusion lighting the command we used was:

{\tt \small \it
tachyon -res 3860 4548 -trans\_vmd -rescale\_lights 0.3 -add\_skylight 1.0 1e79ao.dat
}

\begin{figure}[htbp]
  
   \centerline{\hbox{
   \epsfxsize=2.5in
   \epsffile{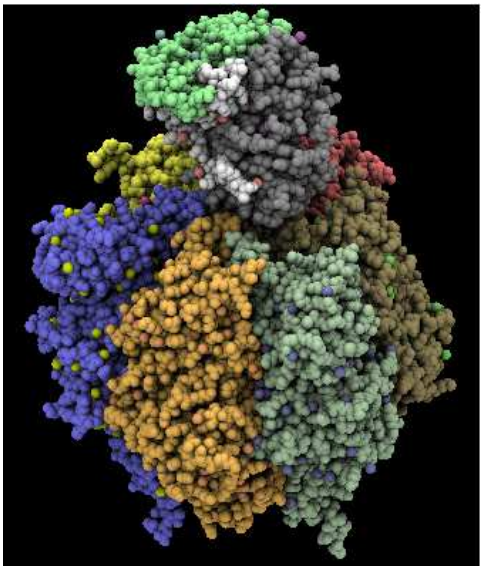}
     }
  }
  \caption{Image of the 1e79 Atp phosphorylase molecule rendered with ambient occlusion lighting on a Blue Gene/L with the parallel tachyon  raytracer. The data consisted  of  26318 objects and was taken from the example file in reference [10].
}
            
  \label{fig1}
  
\end{figure}

The two numbers following the -res command line option indicate the horizontal and vertical number of pixels. The 1e79ao.dat is the input data file from \cite{VMD}.
The command line option -trans\_vmd indicates the presence of ambient occlusion lighting. Ambient occlusion lighting is an  advanced lighting technique that simulates diffuse illumination \cite{VMD}. In conjunction with conventional lighting it can be used to better understand spatial relationships in molecular structures. Since the ambient occlusion lighting mode involves performing Monte Carlo sampling of illumination, rendering with ambient occlusion lighting is computationally intensive. Monte Carlo simulation scales well on the Blue Gene architecture well as gpu architectures \cite{Tomov2} so one expects that dramatic speedups of the Monte Carlo illumination technique should be possible on these systems. Finally probabilistic or Monte Carlo illumination can also be used to simulate realistic outdoor lighting which is related to the graphics Turing scale milestone \cite{Parthenon}. 

High  resolution is required on several visualization systems such as caves or walls and is also an important ingredient in defining the graphics Turing scale. The connection with the graphics Turing scale is that realistic images are processed by the human eye at extremely high resolution. Like advanced lighting, high resolution also adds to the computational difficulty. To illustrate the computational challenge, on a 3GHz Pentium IV computer with 1GB of memory the resolution 1.1 Megapixels rendering with ambient occlusion lighting took  138.9 seconds = 2.315 minutes raytracing time for the rendering of a single frame. For high resolution of 17.6 Megapixels and ambient occlusion lighting the same Pentium IV computer took 2171.6201 seconds = 36.1937 minutes per frame.

\begin{table}[h]
\begin{center}
\begin{tabular}{ |c|c|c|c|} \hline
 Computer   & Ray Tracing Time $t$ & $f= 1/t$& Speedup\\ \hline
Pentium IV &  138.8939 $s$ & 0.00720 $s^{-1}$ & 1 $\times$ \\
64 Processor BGL &  5.6961 $s$ & 0.17556 $s^{-1}$ &  24.384 $\times$ \\
256 Processor BGL &  1.5451 $s$ & 0.64721 $s^{-1}$ & 89.893 $\times$ \\
1024 Processor BGL & .5126 $s$ & 1.95084 $s^{-1}$ & 270.960 $\times$\\ \hline
\end{tabular}
\end{center}
\caption{Data for advanced lighting and a  resolution of $965 \times 1137 $ = 1.1 Megapixels.}
\label{tab2}
\end{table}

\begin{figure}[htbp]
  
   \centerline{\hbox{
   \epsfxsize=5.0in
   \epsffile{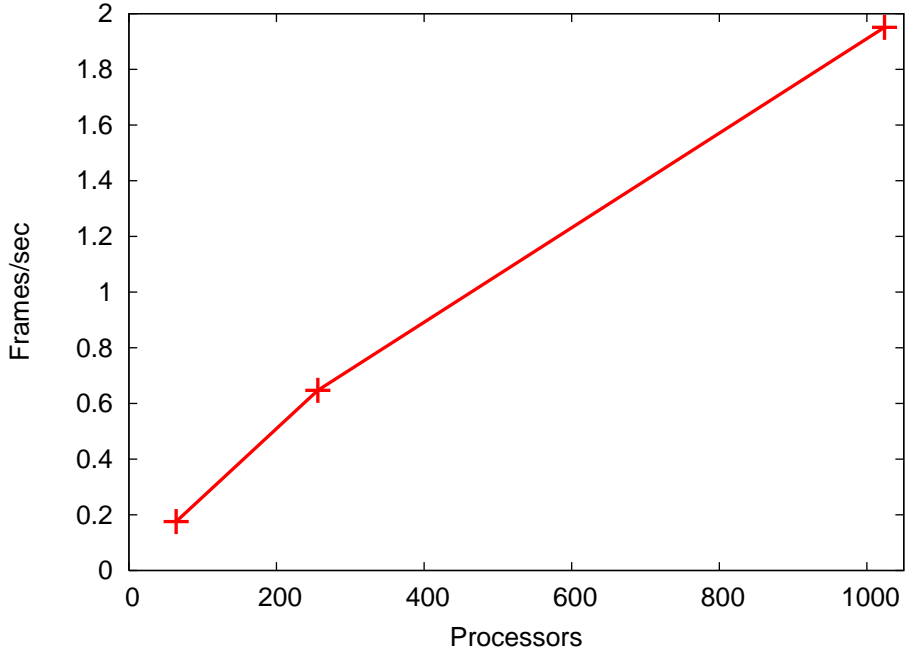}
     }
  }
  \caption{Strong scaling data for a fixed problem size of 1.1 Megapixels or $965 \times 1137$. Perfect strong scaling would be indicated by a linear plot of positive slope.}
            
  \label{fig2}
  
\end{figure}

\begin{table}[h]
\begin{center}
\begin{tabular}{ |c|c|c|c|} \hline
 Computer   & Ray Tracing Time $t$ & $f = 1/t$ & Speedup\\ \hline
Pentium IV&  2171.6201 $s$  & 0.00046 $s^{-1}$ & 1 $\times$ \\
64 Processor BGL &  91.6683 $s$ &  0.01091 $s^{-1}$ &  23.690 $\times$ \\
256 Processor BGL &  24.8193 $s$ & 0.04029 $s^{-1}$ & 87.497 $\times$ \\
1024 Processor BGL & 8.1349 $s$ & 0.12293 $s^{-1}$ & 266.951 $\times$\\
2048 Processor BGL & 5.4128 $s$ & 0.18475 $s^{-1}$ & 401.200 $\times$\\
4096 Processor BGL & 3.3712 $s$ & 0.29663 $s^{-1}$ & 644.168 $\times$\\
6144 Processor BGL & 2.6416 $s$ & 0.37856 $s^{-1}$ & 822.085 $\times$\\ \hline
\end{tabular}
\end{center}
\caption{Data for advanced lighting, resolution of 17.6 Megapixels with $ 3860 \times 4548$ horizontal and vertical number of pixels. Speedup is with respect to a 3GHz Pentium IV.}
\label{tab3}
\end{table}

\begin{figure}[htbp]
  
   \centerline{\hbox{
   \epsfxsize=5.0in
   \epsffile{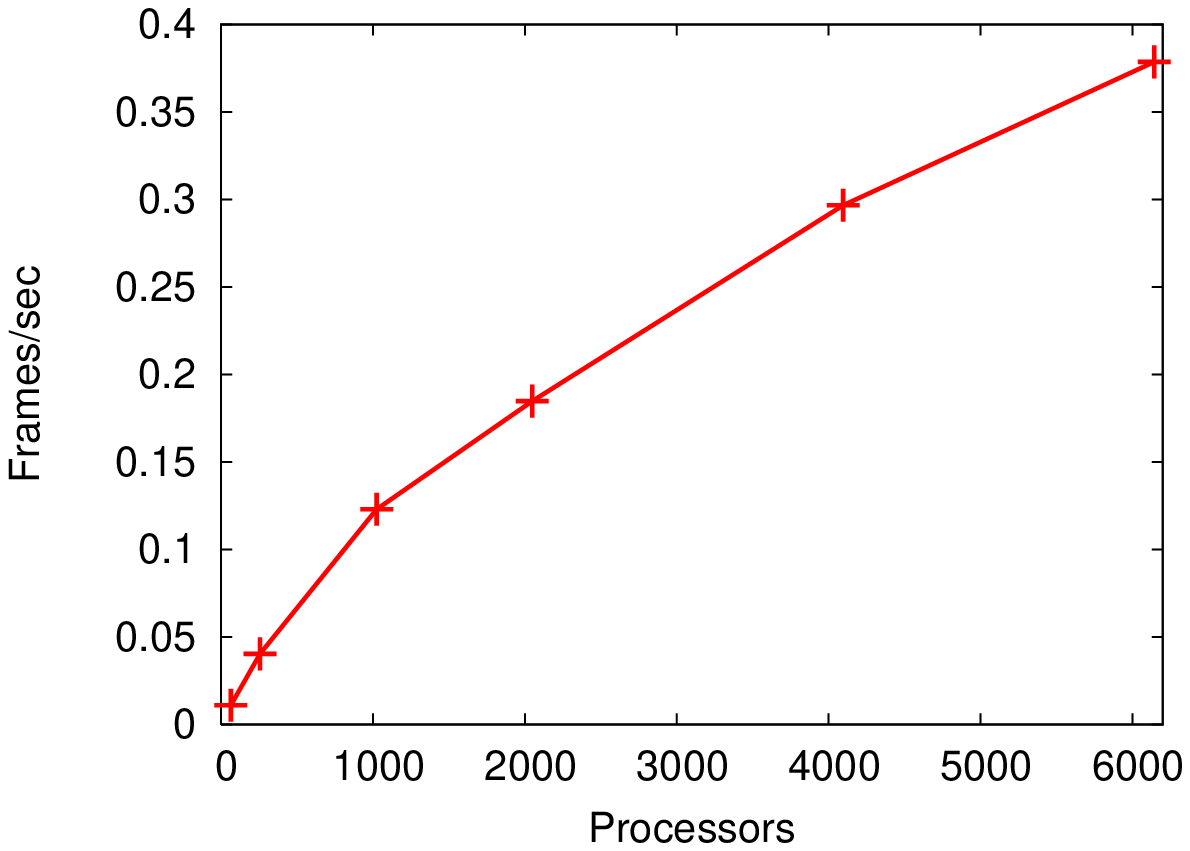}
     }
  }
  \caption{Strong scaling data for a fixed problem size of 17.6 Megapixels with $3860 \times 4548$ horizontal and vertical dimensions. Perfect strong scaling would be indicated by a linear plot of positive slope.}
            
  \label{fig3}
  
\end{figure}

The purpose of the paper is to examine the graphics speedups that are possible using a Blue Gene supercomputer. Table 2 and Table 3 indicate our results for 1.1 Megapixel and 17.6 Megapixel resolution respectively. Our best result was a 822 times speedup at 17.6 Megapixel resolution over the Pentium IV computation. To accomplish that we used a 6144 Processor Blue Gene computer which consisted of 3072 nodes and three racks of the SUNYSB/BNL/NYCCS Blue Gene/L. The variation from a linear plot of Figures 2 and 3 indicate the violation of linear strong scaling at large number of processors. The most likely explanation for this is that communication times between the processors are starting to approach the computation times on each processor when one is using a large number of compute nodes. Thus the processors are spending more time communicating and less time computing and this makes computation less efficient. In the next section we further analyze the performance of the graphics application on the Blue Gene supercomputer.

\section{Weak scaling}

Weak scaling refers increasing the computational workload on a parallel application at the same time as one increases the number of processors. For the raytracing graphics calculation the results of weak scaling are shown in Table 4. To measure weak scaling first we measured the performance on a 64 processor BGL at a resolution of $965 \times 1137$ or 1.1 Megapixels. Then we increased by four the number of processors to 256 at the same time we increased the resolution to $1930 \times 2274$ or 4.4 Megapixels and measured the performance. Finally we increased the number of processors to 1024 at the same time we increased to the resolution to  $3860 \times 4548$ or 17.6 Megapixels.

\begin{table}[h]
\begin{center}
\begin{tabular}{ |c|c|c|c|} \hline
 Computer   & Resolution & Ray Tracing Time $t$ & $f= 1/t$ \\ \hline
64 Processor BGL  & 1.1 Megapixels & 5.6961 $s$ & 0.17556  $s^{-1}$  \\
256 Processor BGL & 4.4 Megapixels &  6.0565 $s$ &     0.16511  $s^{-1}$ \\
1024 Processor BGL & 17.6 Megapixels & 8.1349 $s$ & 0.122927141 $s^{-1}$\\ \hline
\end{tabular}
\end{center}
\caption{Weak scaling data on a Blue Gene/L supercomputer. The 1.1 Megapixel data had  $965 \times 1137$ as the number of pixels in the horizontal and vertical directions. The 4.4 Megapixel data had  $1930 \times 2274$. The 17.6 Megapixel data had  $3860 \times 4548 $ pixels which is more than four times the WQXGA display resolution.}
\label{tab4}
\end{table}

\begin{figure}[htbp]
  
   \centerline{\hbox{
   \epsfxsize=5.0in
   \epsffile{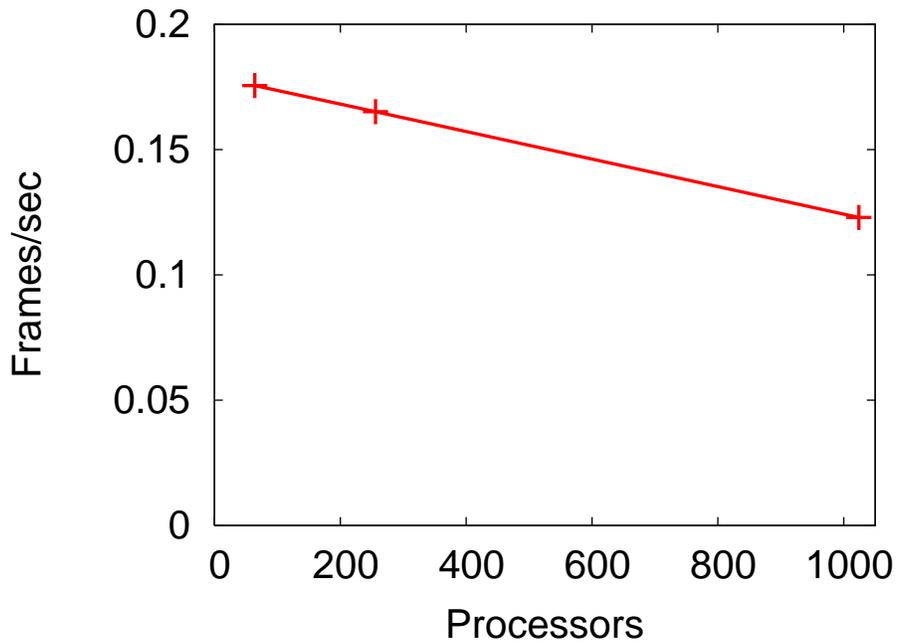}
     }
  }
  \caption{Weak scaling data for varying problem sizes of 1.1 Megapixels, 4.4 Megapixels and 17.6 Megapixels from left to right. Perfect weak  scaling would be indicated by a flat plot of zero slope.}
            
  \label{fig4}
  
\end{figure}

In general raytracing performance is a complicated function of the problem size $N_{pixels}$ and number of processors $N_{processors}$ among other factors. If raytracing performance was a function of the form $f =f(N_{processors}/N_{pixels})$ one would have perfect weak scaling. Note from Figure 4  weak scaling is only good to  thirty percent on 1024 processors which indicates a complicated dependence of the performance of the raytracing calculation on high numbers of processors and problem size. Again the most likely explanation is increased communication between processors at large number of compute nodes. If the raytracing computation obeyed $f \propto  N_{processors}/N_{pixels}$ one would have perfect strong and weak scaling. The results from this and the previous section indicate that this is not the case on the Blue Gene at least for large number of processors.

\section{Conclusion}

Our conclusion is that conventional raytracing software can be run on a Blue Gene and substantial speedups can be obtained. In particular we found a  822 times speedup over a Pentium IV system using a 6144 processor BGL supercomputer. We studied scaling of the raytracing calculation and found complex behavior with deviations from linear scaling at large number of processors.

Although we observed a sizable speedup in ray tracing performance on the Blue Gene supercomputer other aspects of the computation were not improved. For example Table 5 compares the various timings for a high resolution rendering on a Pentium IV and a 6144 processor BGL. Further improvement in scene parsing time and image I/O time will be required before one can approach the graphics Turing scale. One possibility is to read the data into a hierarchical data format which has fast performance characteristics on a Blue Gene computer.

\begin{table}[h]
\begin{center}
\begin{tabular}{ |c|c|c|c|c|} \hline
 Computer   & Scene Parsing& Preprocess& Ray Tracing & Image I/O \\ \hline
Pentium IV&  .8026 $s$  & .1858 $s$ & 2171.6201 $s$ & 1.3193 $s$ \\ \hline
6144 Processor BGL & 5.1100 $s$ & .1868 $s$  & 2.6416 $s$ & 8.5864 $s$ \\ \hline
\end{tabular}
\end{center}
\caption{Detailed data for advanced lighting and high resolution of 17.6 Megapixels.}
\label{tab5}
\end{table}

Another point is that although the Blue Gene we used is among the world's fastest computers it is not the fastest. Table 6 lists the characteristics of the world's fastest computer at this time, the DOE/NNSA/LLNL facility. The 6144 processor Blue Gene that we used to achieve a 822 times speedup over a Pentium IV represents less than three percent of the largest supercomputer's capacity.  It would be interesting to measure the raytracing performance that can be achieved on such a system.

\begin{table}[h]
\begin{center}
\begin{tabular}{ |c|c|c|c|c|c|c|} \hline
 Rank   & Facility & Racks & Nodes & Processors & Peak Speed & Memory\\ \hline
1 & LLNL & 104 racks & 106496 & 212992 &  596 Teraflop/s & 73.7 TB\\ \hline
\end{tabular}
\end{center}
\caption{Characteristics of the of the worlds fastest Blue Gene supercomputer. $1$ Teraflop/s $= 10^{12}$ floating point operations per second. $1$ TB $=1000$ Gigabytes. }
\label{tab6}
\end{table}

Finally although our approach to parallel raytracing on the Blue Gene/L supercomputer can be considered a brute force approach compared to gpu programming or realtime raytracing hardware, it should not be surprising that such a solution could achieve this computational milestone. Other computational milestones such as defeating expert chess players, or computing atomic structure were first done on large supercomputers. Advances in software, algorithms and hardware allows the same tasks to be performed with more modest resources today. The largest supercomputers will then set their sights on even more ambitious computational challenges. 

\section*{Acknowledgments}
This manuscript has been authored in part by Brookhaven Science Associates, LLC, under Contract No. DE-AC02-98CH10886 with the U.S. Department of Energy. We wish to thank Len Slatest, Nick D'Imperio and Stratos Efstathiadis for valuable assistance.

\end{document}